\documentclass[preprint,12pt]{elsarticle}
\usepackage{natbib}
\usepackage{graphicx}
\usepackage{amsmath,amssymb}
\usepackage{float}
\usepackage{color}
\usepackage[basic,box]{circ}

\newcommand{\pref}[1]{(\ref{#1})}

\newcommand{\Eq}[1]{Eq.~(\ref{#1})}
\newcommand{\Fig}[1]{Fig.~\ref{#1}}

\newcommand{\Tab}[1]{Tab.~\ref{#1}}

\newcommand{\Df}{\mathrm{d}}
\newcommand{\Ex}{\mathrm{e}}
\newcommand{\quant}[2]{#1 \, \mathrm{#2}}

\bibpunct{(}{)}{,}{}{}{,}

\journal{Frontiers in Computational Neuroscience}

\begin{document}

\bibliographystyle{chicago}

\begin{frontmatter}

\title{\textbf{A biophysical observation model for field potentials of networks of leaky integrate-and-fire neurons}}

\author{Peter beim Graben\corref{cor1}}
\address{Bernstein Center for Computational Neuroscience Berlin, Germany}
\address{Department of German Language and Linguistics,
 Humboldt-Universit\"at zu Berlin}
\ead{peter.beim.graben@hu-berlin.de}
\ead[url]{www.beimgraben.info}
\cortext[cor1]{Department of German Language and Linguistics \\
 Humboldt-Universit\"at zu Berlin \\
 Unter den Linden 6 \\
 D -- 10099 Berlin \\
 Phone: +49-30-2093-9632 \\
 Fax: +49-30-2093-9729
 }

\author{Serafim Rodrigues\corref{nc}}
\address{Centre for Robotics and Neural Systems, School of Computing and Mathematics, \\
University of Plymouth, United Kingdom}

\begin{abstract}
We present a biophysical approach for the coupling of neural network activity as resulting from proper dipole currents of cortical pyramidal neurons to the electric field in extracellular fluid. Starting from a reduced three-compartment model of a single pyramidal neuron, we derive an observation model for dendritic dipole currents in extracellular space and thereby for the dendritic field potential that contributes to the local field potential of a  neural population. This work aligns and satisfies the widespread dipole assumption that is motivated by the ``open-field'' configuration of the dendritic field potential around cortical pyramidal cells. Our reduced three-compartment scheme allows to derive networks of leaky integrate-and-fire models, which facilitates comparison with existing neural network and observation models. In particular, by means of numerical simulations we compare our approach with an \emph{ad hoc} model by Mazzoni et al. [Mazzoni, A., S.~Panzeri, N.~K. Logothetis, and N.~Brunel (2008). Encoding of naturalistic stimuli by local field potential spectra in networks of excitatory and inhibitory neurons. {\em PLoS Computational Biology\/}~{\em 4\/}(12), e1000239], and conclude that our biophysically motivated approach yields substantial improvement.
\end{abstract}


\end{frontmatter}

\section{Introduction}
\label{sec:intro}

\begin{figure}[H]
 \centering
 \includegraphics[width=1\textwidth]{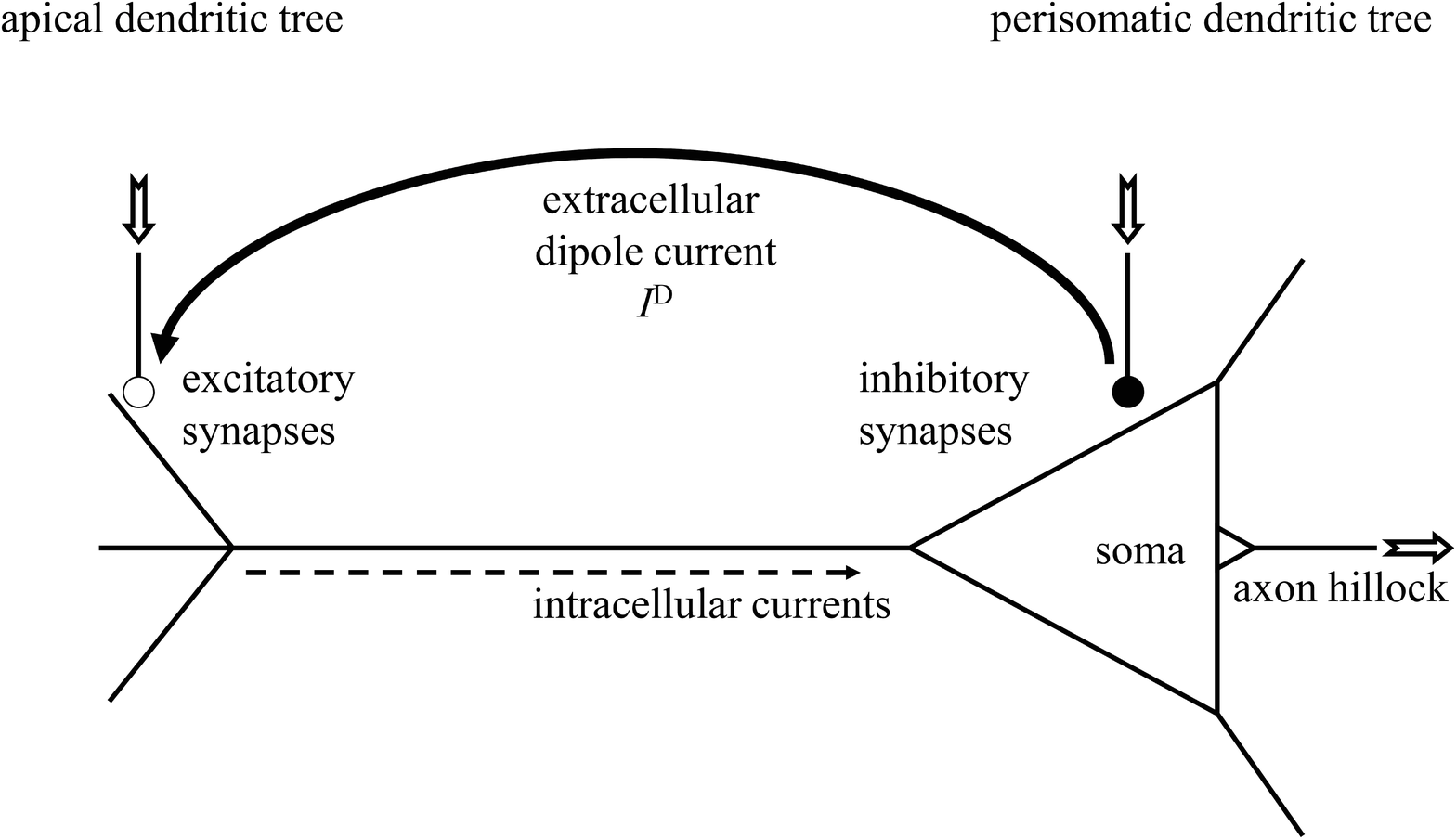}
\caption{\label{fig:pyramid} Sketch of a cortical pyramidal neuron with extracellular current dipole between spatially separated excitatory (open bullet) and inhibitory synapses (filled bullet). Neural in- and outputs are indicated by the jagged arrows. Dendritic current $I^\mathrm{D}$ causes dendritic field potential (DFP).}
\end{figure}

Since Hans Berger's 1924 discovery of the human \emph{electroencephalogram (EEG)} \citep{Berger29}, neuroscientists achieved much progress in clarifying its neural generators \citep{CreutzfeldtWatanabeLux66a, CreutzfeldtWatanabeLux66b, NunezSrinivasan06, SchomerSilva11}. These are the cortical pyramidal neurons, as sketched in \Fig{fig:pyramid}, that possess a long dendritic trunk separating mainly excitatory synapses at the apical dendritic tree from mainly inhibitory synapses at the soma and at the perisomatic basal dendritic tree \citep{CreutzfeldtWatanabeLux66a, Spruston08}. In addition, they exhibit an axial symmetry and are aligned in parallel to each other, perpendicular to the cortex' surface, thus forming a palisade of cell bodies and dendritic trunks. When both kinds of synapses are simultaneously active, inhibitory synapses generate current sources and excitatory synapses current sinks in extracellular space, hence causing the pyramidal cell to behave as a microscopic dipole surrounded by its characteristic electrical field, the \emph{dendritic field potential (DFP)}. The densely packed pyramidal cells form then a dipole layer whose superimposed currents give rise to the \emph{local field potential (LFP)} of neural masses and eventually to the EEG \citep{LindenPettersenEinevoll10, LindenTetzlaffEA11, NunezSrinivasan06, SchomerSilva11}.

Despite of the progress from experimental neuroscience, theoretically understanding the coupling of complex neural network dynamics to the electromagnetic field in the extracellular space poses challenging problems; some of them have been addressed to some extent by \citet{BedardKroegerDestexhe04, BedardDestexhe09}, and \citet{BedardDestexhe12}.

In computer simulation studies, neural mass potentials, such as LFP and EEG are most realistically simulated by means of multicompartmental models \citep{LindenPettersenEinevoll10, LindenTetzlaffEA11, ProtopapasVanierBower98, SargsyanPapatheodoropoulosKostopoulos01}. \Citet{LindenPettersenEinevoll10} calculated the current dipole momentum of the DFP for single pyramidal and stellate cells, based on several hundreds compartments of the dendritic trees. Their results were in compliance with the standard dipole approximation of the electrostatic multipole expansion in the far-field (more than 1 mm remote from the dendritic trunk), but they found rather poor agreement with that approximation in the vicinity of the cell body. For comparison they also computed a ``two-monopole'' model of one synaptic current and its counterpart, the somatic return current, estimated from the current dipole momentum of the whole dendritic tree. This ``two-monopole'' model, which corresponds to an electrically equivalent single dipole model, obtained from the decomposition of the dendrite into two compartments, better approximates the true current dipole momentum in the vicinity of the pyramidal neuron. By superimposing the DFPs of pyramidal cells to the ensemble LFP, \citet{LindenTetzlaffEA11} found that LFP properties cannot be attributed to the far-field dipole approximation.

However, realistic multicompartmental models are computationally too expensive for large-scale neural network simulations. Therefore, various techniques have been proposed and employed to overcome computational complexity. These include networks of point models (i.e. devoid from any spatial representation), based on conductance models \citep{HodgkinHuxley52, MazzoniPanzeriEA08}, population density models \citep{omurtag2000simulation}, or firing rate models~\citep{WilsonCowan72}, which can be seen as a sub class of population density models, with uniform density distribution \citep{Chizhov2007}. In these kinds of models, mass potentials such as LFP or EEG are conventionally described as averaged membrane potential. A different class of models are neural mass models \citep{DavidFriston03, JansenRit95, WendlingBellangerEA00, rodrigues2010mappings}, where mass potentials are estimated either through sums (or actually differences) of postsynaptic potentials \citep{DavidFriston03} or of postsynaptic currents \citep{MazzoniPanzeriEA08}.

In particular, the model of \citet{MazzoniPanzeriEA08} which is based on \citet{BrunelWang03}, recently led to a series of follow-up studies \citep{MazzoniEA10, MazzoniBrunelEA11} addressing the correlations between numerically simulated and experimentally measured LFP/EEG with spike rates by means of statistical modeling and information theoretic measures. In all of the above point models and their extension to population models, it is assumed that the extracellular space is iso-potential and the majority of studies thereby neglect the effect of extracellular resistance. That is, the extracellular space constitutes a different and isolated domain with no effect on neuronal dynamics.

In this article we extend the \emph{ad hoc} model of \citet{MazzoniPanzeriEA08} towards a biophysically better justified approach, taking the dipole character of extracellular currents and fields into account. Basically, our model corresponds to the ``two-monopole'', or, equivalent dipole model of \citet{LindenPettersenEinevoll10} which gave a good fit of the DFP close to the cell body of a cortical pyramidal neuron. However, we aim to keep the simplicity of the \citet{MazzoniPanzeriEA08} model in terms of computational complexity, by endowing the extracellular space with resistance and by keeping point-like neuronal circuits. That is, in our case we do not quite consider point neurons, nor spatially extended models with detailed compartmental morphology, yet an intermediate level of description is achieved. To this end we propose a \emph{reduced three-compartmental model} of a single pyramidal neuron \citep{Destexhe01, Graben08a, WangTegnerEA04}, and derive an observation model for the dendritic dipole currents in the extracellular space and thereby for the DFP that contributes to the LFP of a neural population. Interestingly, our reduced three-compartmental model enables us to derive a leaky integrate-and-fire mechanism (as for a point model \citep{MazzoniPanzeriEA08}), with additional observation equations for the DFP, which all together allows to study the relationship between spike rates and LFP. Our derivations also nicely map realistic electrotonic parameters to phenomenological parameters considered in \citet{MazzoniPanzeriEA08}.

\section{Material and Methods}
\label{sec:meth}

\Citet{MazzoniPanzeriEA08} consider three populations of neurons, namely excitatory cortical pyramidal cells (population 1), inhibitory cortical interneurons (population 2) and excitatory thalamic relay neurons (population 3), passing sensory input to the cortex that is simulated by a random (Erd\H{o}s-R\'{e}nyi) graph of $K = 4000$ pyramidal and $L = 1000$ interneurons with connection probability $P = 0.2$.

\subsection{Theory}
\label{sec:theo}

\begin{figure}[H]
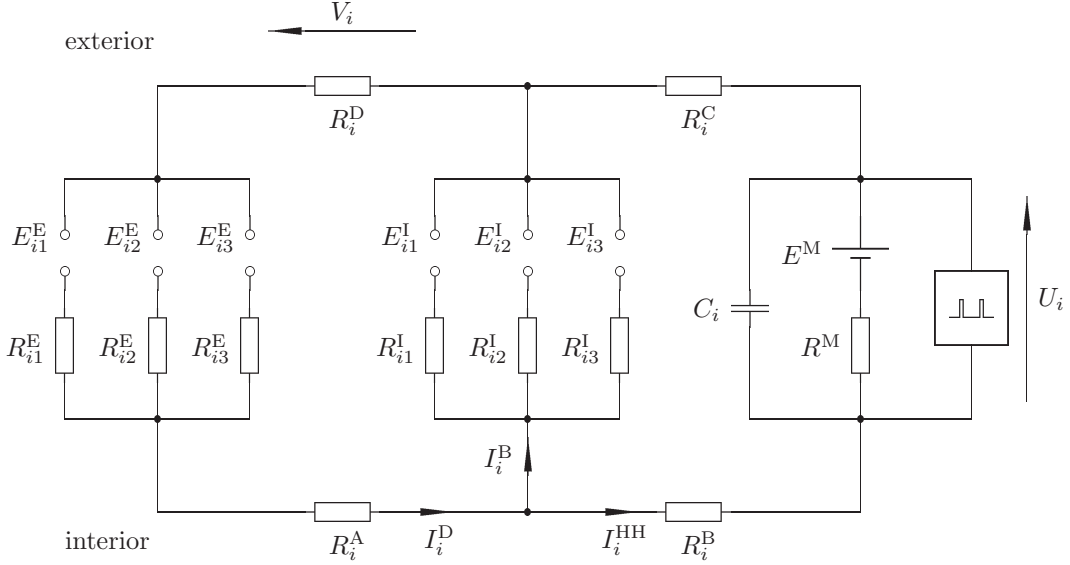

 \centering
 \begin{circuit}0
 \.1
 \- 5 d
 \.2
 \- 2 d
 \connection1 { $E^\mathrm{I}_{i2}$} c d
 \_ 2 d
 \connection2 {} c u
 \_ 2 d
 \- 1 d
 \nl \R1 { $R^\mathrm{I}_{i2}$} d
 \- 2 d
 \.3
 \- 5 d
 \.4
 \atpin .2
 \- 5 l
 \- 2 d
 \connection3 { $E^\mathrm{I}_{i1}$} c d
 \_ 2 d
 \connection4 {} c u
 \_ 2 d
 \- 1 d
 \nl \R2 { $R^\mathrm{I}_{i1}$} d
 \- 2 d
 \- 5 r
 \atpin .2
 \- 5 r
 \- 2 d
 \connection5 { $E^\mathrm{I}_{i3}$} c d
 \_ 2 d
 \connection6 {} c u
 \_ 2 d
 \- 1 d
 \nl \R3 { $R^\mathrm{I}_{i3}$} d
 \- 2 d
 \- 5 l
 \atpin .1
 \- 7 r
 \nl \R10 { $R^\mathrm{C}_i$} r
 \- 7 r
 \- 5 d
 \.5
 \- 2 d
 \nl \U1 { $E^\mathrm{M}$} + d
 \- 1 d
 \nl \R4 { $R^\mathrm{M}$} d
 \- 2 d
 \.6
 \- 5 d
 \atpin .5
 \- 6 l
 \- 6 d
 \nl \C1 { $C_i$} d
 \- 5 d
 \- 6 r
 \atpin .5
 \- 6 r
 \vcenterto C1
 \Impulse1 P4 {} {} {} {}
 \frompin Impulse1P4
 \- 5 u
 \frompin Impulse1P2
 \- 4 d
 \- 6 l
 \atpin .1
 \- 8 l
 \nl \R5 { $R^\mathrm{D}_i$} l
 \- 8 l
 \- 5 d
 \.7
 \- 2 d
 \connection7 { $E^\mathrm{E}_{i2}$} c d
 \_ 2 d
 \connection8 {} c u
 \_ 2 d
 \- 1 d
 \nl \R6 { $R^\mathrm{E}_{i2}$} d
 \- 2 d
 \.8
 \- 5 d
 \atpin .7
 \- 5 l
 \- 2 d
 \connection9 { $E^\mathrm{E}_{i1}$} c d
 \_ 2 d
 \connection10 {} c u
 \_ 2 d
 \- 1 d
 \nl \R7 { $R^\mathrm{E}_{i1}$} d
 \- 2 d
 \- 5 r
 \atpin .7
 \- 5 r
 \- 2 d
 \connection11 { $E^\mathrm{E}_{i3}$} c d
 \_ 2 d
 \connection12 {} c u
 \_ 2 d
 \- 1 d
 \nl \R8 { $R^\mathrm{E}_{i3}$} d
 \- 2 d
 \- 5 l
 \atpin .8
 \- 5 d
 \- 8 r
 \nl \R11 { $R^\mathrm{A}_i$} r
 \- 8 r
 \atpin .4
 \- 7 r
 \nl \R12 { $R^\mathrm{B}_i$} r
 \- 7 r
 \atpin .8
 \moverel{16} {-5}
 \nl \whatI2 { $I^\mathrm{D}_i$} d l
 \atpin .4
 \moverel{0} {2}
 \nl \whatI3 { $I^\mathrm{B}_i$} s u
 \atpin .4
 \moverel{6} {0}
 \nl \whatI4 { $I^\mathrm{HH}_i$} d l
 \atpin .2: \shift -5 8 \P1
 \atpin .7: \shift 5 8 \P2
 \Utext {$V_i$} from P1 to P2
 \atpin .5: \shift 9 0 \P3
 \atpin .6: \shift 9 0 \P4
 \Utext {$U_i$} from P4 to P3
 \atpin .7: \shift -6 14
 \put{apical dendritic tree}
 \atpin .2: \shift -6 14
 \put{perisomatic dendritic tree}
 \atpin .5: \shift -5 14
 \put{axon hillock}
 \atpin .7: \shift -5 7
 \put{exterior}
 \atpin .8: \shift -5 -7
 \put{interior}
\end{circuit}
\caption{\label{fig:eqcirc} Proposed electronic equivalent circuit for a pyramidal neuron (reduced three compartmental model). Note that the apical and basal dendrites are not true compartments since capacitors are not explicitly represented, rather, these are implicitly taken into account via EPSP and IPSP static functions, thus keeping computational complexity low.}
\end{figure}

We describe the $i$th cortical pyramidal neuron [\Fig{fig:pyramid}] from population 1 via the electronic equivalent (reduced) three-compartment model \Fig{fig:eqcirc} \citep{Destexhe01, Graben08a, WangTegnerEA04}, which is parsimonious to derive our observation model: one compartment for the apical dendritic tree, another one for soma and perisomatic basal dendritic tree \citep{LindenPettersenEinevoll10}, and the third --- actually a leaky integrate-and-fire (LIF) unit --- for the axon hillock where membrane potential is converted into spike trains by means of an integrate-and-fire mechanism.

Excitatory synapses are represented by the left-most branch, where excitatory postsynaptic potentials (EPSP) at a synapse between a neuron $j$ from population 1 or 3 and neuron $i$ act as electromotoric forces $E^\mathrm{E}_{ij}$. These potentials drive excitatory postsynaptic currents (EPSC) $I^\mathrm{E}_{ij}$, essentially consisting of sodium ions, through the cell plasma with resistance $R^\mathrm{E}_{ij}$ from the synapse towards the axon hillock.

The middle branch describes the inhibitory synapses between a neuron $k$ from population 2 and neuron $i$. Here, inhibitory postsynaptic potentials (IPSP) $E^\mathrm{I}_{ik}$ provide a shortcut between the excitatory branch and the trigger zone, where inhibitory postsynaptic currents (IPSC) $I^\mathrm{I}_{ik}$ (essentially chloride ions) close the loop between the apical and perisomatic dendritic trees. The resistivity of the current paths along the cell plasma is given by $R^\mathrm{I}_{ik}$.

The cell membrane at the axon hillock itself is represented by the branch at the right hand side. Here, a capacitor $C_i$ reflects the temporary storage capacity of the membrane. The serial circuit consisting of a battery $E^\mathrm{M}$ and a resistor $R^\mathrm{M}$ denotes the Nernst resting potential and the leakage conductance of the membrane, respectively \citep{JohnstonWu97}. Finally, a spike generator \citep{HodgkinHuxley52, MazzoniPanzeriEA08} (indicated by a ``black box'') is regarded of having infinite input impedance. Both, EPSP and IPSP result from the interaction of postsynaptic receptor kinetics with dendritic low-pass filtering in compartments one and two, respectively \citep{DestexheMainenSejnowski98, LindenPettersenEinevoll10}. Hence the required capacitances, omitted in \Fig{fig:eqcirc}, are already taken into account by $E^\mathrm{E}_{ij}, E^\mathrm{I}_{ik}$. Therefore, we refer to our model as to a ``reduced compartment model'' here.

The three compartments are coupled through longitudinal resistors, $R^\mathrm{A}_i, R^\mathrm{B}_i, R^\mathrm{C}_i, R^\mathrm{D}_i$ where $R^\mathrm{A}_i, R^\mathrm{B}_i$ denote the resistivity of the cell plasma and $R^\mathrm{C}_i, R^\mathrm{D}_i$ that of extracellular space \citep{HoltKoch99a}.

Finally, the membrane voltage at the axon hillock $U_i$ (the dynamical state variable) and the DFP $V_i$, which measures the drop in electrical potential along the extracellular resistor $R^\mathrm{D}_i$ are indicated. For the aim of calculation, the mesh currents $I^\mathrm{D}_i$ (the dendritic current), $I^\mathrm{B}_i$ (the basal current) and $I^\mathrm{IF}_i$ (the integrate-and-fire current) are indicated.

The circuit in \Fig{fig:eqcirc} obeys the following equations:
\begin{eqnarray}
 \label{eq:circuit1}
  I^\mathrm{D}_i &=& \sum_{j = 1}^p I^\mathrm{E}_{ij} \\
 \label{eq:circuit2}
  I^\mathrm{B}_i &=& \sum_{k = 1}^q I^\mathrm{I}_{ik} \\
 \label{eq:circuit3}
  I^\mathrm{IF}_i &=& I^\mathrm{D}_i - I^\mathrm{B}_i \\
 \label{eq:circuit4}
  I^\mathrm{IF}_i &=&  C_i \frac{\Df U_i}{\Df t} + \frac{U_i - E^\mathrm{M}}{R^\mathrm{M}} \\
 \label{eq:circuit5}
 E^\mathrm{E}_{ij} &=& R^\mathrm{E}_{ij} I^\mathrm{E}_{ij} +
     ( R^\mathrm{A}_i + R^\mathrm{D}_i ) I^\mathrm{D}_i +
     ( R^\mathrm{B}_i + R^\mathrm{C}_i ) I^\mathrm{IF}_i + U_i \:, 1 \le j \le p \\
 \label{eq:circuit6}
 E^\mathrm{I}_{ik} &=& R^\mathrm{I}_{ik} I^\mathrm{I}_{ik} +
        ( R^\mathrm{B}_i + R^\mathrm{C}_i ) I^\mathrm{IF}_i + U_i \:, 1 \le k \le q \\
 \label{eq:circuit7}
  V_i &=& R^\mathrm{D}_i I^\mathrm{D}_i \:.
\end{eqnarray}
Here, $p$ is the number of excitatory and $q$ is the number of inhibitory synapses connected to neuron $i$.

The circuit described by Eqs. (\ref{eq:circuit1} -- \ref{eq:circuit7}) shows that the neuron $i$ is likely to fire when the excitatory synapses are activated. Then, the integrate-and-fire current $I^\mathrm{IF}_i$ equals the dendritic current $I^\mathrm{D}_i$. If, by contrast, also the inhibitory synapses are active, the dendritic current $I^\mathrm{D}_i$ is shunted between the apical and perisomatic basal dendritic trees and only a portion could evoke spikes at the trigger zone [\Eq{eq:circuit4}]. On the other hand, the large dendritic current $I^\mathrm{D}_i$ flowing through the extracellular space of resistance $R^\mathrm{D}_i$, gives rise to a large DFP $V_i$.

In order to simplify the following derivations, we gauge the resting potential [\Eq{eq:circuit4}] to $E^\mathrm{M} = 0$, yielding
\begin{equation}
\label{eq:hhsimple1}
 I^\mathrm{IF}_i =  C_i \frac{\Df U_i}{\Df t} + \frac{U_i}{R^\mathrm{M}} \:.
\end{equation}

From \pref{eq:circuit5} we obtain the individual EPSC's as
\begin{equation}
\label{eq:epsc}
  I^\mathrm{E}_{ij} =
  \frac{1}{R^\mathrm{E}_{ij}} \left[ E^\mathrm{E}_{ij} -
    (R^\mathrm{A}_i + R^\mathrm{D}_i) I^\mathrm{D}_i  -
    (R^\mathrm{B}_i + R^\mathrm{C}_i) I^\mathrm{IF}_i - U_i
  \right] \:.
\end{equation}

And accordingly, the individual IPSC's from \pref{eq:circuit6}
\begin{equation}
\label{eq:ipsc}
  I^\mathrm{I}_{ik} =
  \frac{1}{R^\mathrm{I}_{ik}} \left[ E^\mathrm{I}_{ik} -
    (R^\mathrm{B}_i + R^\mathrm{C}_i) I^\mathrm{IF}_i - U_i
  \right] \:.
\end{equation}

Inserting \pref{eq:epsc} into \pref{eq:circuit1} yields the excitatory dendritic current
\begin{equation}
\label{eq:edc}
 I^\mathrm{D}_i = \sum_{j = 1}^p  \frac{1}{R^\mathrm{E}_{ij}} E^\mathrm{E}_{ij} \, - \,
 g^\mathrm{E}_i [ (R^\mathrm{A}_i + R^\mathrm{D}_i) I^\mathrm{D}_i  +
    (R^\mathrm{B}_i + R^\mathrm{C}_i) I^\mathrm{IF}_i + U_i ] \:,
\end{equation}
where we have introduced the excitatory dendritic conductivity
\begin{equation}
\label{eq:exconduct}
 g^\mathrm{E}_i = \sum_{j = 1}^p  \frac{1}{R^\mathrm{E}_{ij}} \:.
\end{equation}

Likewise we obtain the inhibitory dendritic currents from \pref{eq:circuit2} and
\pref{eq:ipsc} as
\begin{equation}
\label{eq:idc}
 I^\mathrm{B}_i  = \sum_{k = 1}^q  \frac{1}{R^\mathrm{I}_{ik}} E^\mathrm{I}_{ik} \, - \,
 g^\mathrm{I}_i [ (R^\mathrm{B}_i + R^\mathrm{C}_i) I^\mathrm{IF}_i + U_i ] \:,
\end{equation}
with the inhibitory dendritic conductivity
\begin{equation}
\label{eq:inconduct}
 g^\mathrm{I}_i = \sum_{k = 1}^q  \frac{1}{R^\mathrm{I}_{ik}} \:.
\end{equation}

With these results, we obtain an interface equation for an observation model as follows. Rearranging \pref{eq:edc} yields
\begin{equation}
 \label{eq:id}
 I^\mathrm{D}_i [ 1 + g^\mathrm{E}_i (R^\mathrm{A}_i + R^\mathrm{D}_i) ] =
     \sum_{j = 1}^p  \frac{1}{R^\mathrm{E}_{ij}} E^\mathrm{E}_{ij} \, - \,
    g^\mathrm{E}_i [ (R^\mathrm{B}_i + R^\mathrm{C}_i) I^\mathrm{IF}_i + U_i ]
\end{equation}

Next, we eliminate $I^\mathrm{IF}_i$ through \pref{eq:hhsimple1}:
\begin{eqnarray*}
 I^\mathrm{D}_i \left[ 1 + g^\mathrm{E}_i \left( R^\mathrm{A}_i + R^\mathrm{D}_i \right) \right] &=&
     \sum_{j = 1}^p  \frac{1}{R^\mathrm{E}_{ij}} E^\mathrm{E}_{ij} \, - g^\mathrm{E}_i \left[ C_i \left( R^\mathrm{B}_i + R^\mathrm{C}_i \right)
        \frac{\Df U_i}{\Df t} + U_i \left( 1 + \frac{R^\mathrm{B}_i + R^\mathrm{C}_i}{R^\mathrm{M}} \right)
        \right] \:.
\end{eqnarray*}

Division by $1 + g^\mathrm{E}_i \left( R^\mathrm{A}_i + R^\mathrm{D}_i \right)$ gives the desired expression for the extracellular dendritic dipole current:
\begin{equation}
 \label{eq:dfc}
    I^\mathrm{D}_i  = \sum_{j = 1}^p  \alpha_{ij} E^\mathrm{E}_{ij} -
    \beta_i \frac{\Df U_i}{\Df t} - \gamma_i U_i  \:,
\end{equation}
with the following electrotonic parameters
\begin{eqnarray}
  \label{eq:alpha} \alpha_{ij} &=& \frac{1}{R^\mathrm{E}_{ij}[ 1 + g^\mathrm{E}_i (R^\mathrm{A}_i + R^\mathrm{D}_i) ]} \\
  \label{eq:beta} \beta_i &=& \frac{C_i g^\mathrm{E}_i (R^\mathrm{B}_i + R^\mathrm{C}_i)}{
 1 + g^\mathrm{E}_i (R^\mathrm{A}_i + R^\mathrm{D}_i)} \\
  \label{eq:gamma} \gamma_i &=& \frac{g^\mathrm{E}_i (R^\mathrm{M} + R^\mathrm{B}_i + R^\mathrm{C}_i)}{
        R^\mathrm{M} [ 1 + g^\mathrm{E}_i (R^\mathrm{A}_i + R^\mathrm{D}_i)]} \:.
\end{eqnarray}

In order to derive the evolution equation we consider the integrate-and-fire current $I^\mathrm{IF}_i$ that is given through \pref{eq:circuit3}. The individual EPSCs and IPSCs have already been obtained in \pref{eq:epsc} and \pref{eq:ipsc}, respectively. Inserting \pref{eq:idc} into \pref{eq:circuit3} yields

\begin{eqnarray*}%
 I^\mathrm{IF}_i [1 -  g^\mathrm{I}_i (R^\mathrm{B}_i + R^\mathrm{C}_i) ] -  g^\mathrm{I}_i U_i
 &=& I^\mathrm{D}_i - \sum_{k = 1}^q \frac{1}{R^\mathrm{I}_{ik}} E^\mathrm{I}_{ik} \:.
\end{eqnarray*}

Next we insert our interface equation \pref{eq:dfc} and also \pref{eq:hhsimple1}:
\begin{eqnarray*}
 \left[ C_i \frac{\Df U_i}{\Df t} + \frac{U_i}{R^\mathrm{M}} \right]
    [1 -  g^\mathrm{I}_i (R^\mathrm{B}_i + R^\mathrm{C}_i) ] -  g^\mathrm{I}_i U_i
 &=& \sum_{j = 1}^p  \alpha_{ij} E^\mathrm{E}_{ij} -
    \beta_i \frac{\Df U_i}{\Df t} - \gamma_i U_i -
     \sum_{k = 1}^q \frac{1}{R^\mathrm{I}_{ik}} E^\mathrm{I}_{ik}
\end{eqnarray*}
and obtain after some rearrangements
\begin{eqnarray*}
 \{ C_i \left[ 1 -  g^\mathrm{I}_i (R^\mathrm{B}_i + R^\mathrm{C}_i) \right] + \beta_i  \} \frac{\Df U_i}{\Df t}
 &+&
 \frac{1 - g^\mathrm{I}_i (R^\mathrm{B}_i + R^\mathrm{C}_i + R^\mathrm{M}) + R^\mathrm{M} \gamma_i}{R^\mathrm{M}} U_i = \\
 && \sum_{j = 1}^p  \alpha_{ij} E^\mathrm{E}_{ij} -
 \sum_{k = 1}^q \frac{1}{R^\mathrm{I}_{ik}} E^\mathrm{I}_{ik}
\end{eqnarray*}
and after multiplication with
\[
r_i = \frac{R^\mathrm{M}}{1 - g^\mathrm{I}_i (R^\mathrm{B}_i + R^\mathrm{C}_i + R^\mathrm{M})  + R^\mathrm{M} \gamma_i}
\]
the dynamical law for the membrane potential at axon hillock:
\begin{equation}
 \label{eq:lifneuron}
 \tau_i \frac{\Df U_i}{\Df t} + U_i =
 \sum_{j = 1}^p w^\mathrm{E}_{ij} \, E^\mathrm{E}_{ij} -
 \sum_{k = 1}^q w^\mathrm{I}_{ik} \, E^\mathrm{I}_{ik} \:,
\end{equation}
where we have introduced the following parameters:
\begin{itemize}
    \item \emph{time constants}
    \begin{equation}
    \label{eq:timeconst}
      \tau_i = r_i \{ C_i \left[ 1 -  g^\mathrm{I}_i (R^\mathrm{B}_i + R^\mathrm{C}_i) \right] + \beta_i  \}
    \end{equation}
    \item \emph{excitatory synaptic weights}
    \begin{equation}
    \label{eq:exweig}
        w^\mathrm{E}_{ij} = r_i \alpha_{ij}
    \end{equation}
    \item \emph{inhibitory synaptic weights}
    \begin{equation}
    \label{eq:inweig}
        w^\mathrm{I}_{ik} = \frac{r_i}{R^\mathrm{I}_{ik}} \:.
    \end{equation}
\end{itemize}

Using the result \pref{eq:lifneuron}, we can also eliminate the temporal derivative in the interface equation \pref{eq:dfc} through
\begin{equation}
\label{eq:dudt}
 \frac{\Df U_i}{\Df t} =
 \frac{1}{\tau_i} \left[
     \sum_{j = 1}^p w^\mathrm{E}_{ij} \, E^\mathrm{E}_{ij} -
     \sum_{k = 1}^q w^\mathrm{I}_{ik} \, E^\mathrm{I}_{ik} - U_i
 \right]
\end{equation}
which yields
\begin{eqnarray*}
  I^\mathrm{D}_i &=&
 \sum_{j = 1}^p \left( \alpha_{ij} - \frac{\beta_i}{\tau_i} w^\mathrm{E}_{ij} \right) \,
 E^\mathrm{E}_{ij} + \sum_{k = 1}^q \frac{\beta_i}{\tau_i} w^\mathrm{I}_{ik} \, E^\mathrm{I}_{ik} +
  \left( \frac{\beta_i}{\tau_i} - \gamma_i \right) U_i \:.
\end{eqnarray*}
And eventually, by virtue of \Eq{eq:circuit7} after multiplication with $R^\mathrm{D}_i$ the dendritic field potential (DFP)
\begin{equation}
 \label{eq:dfc2}
  V_i =
 \sum_{j = 1}^p \tilde{w}^\mathrm{E}_{ij} \, E^\mathrm{E}_{ij} +
 \sum_{k = 1}^q \tilde{w}^\mathrm{I}_{ik} \, E^\mathrm{I}_{ik} +
 \xi_i U_i \:,
\end{equation}
with parameters
\begin{eqnarray}
 \label{eq:exti} \tilde{w}^\mathrm{E}_{ij} &=& R^\mathrm{D}_i w^\mathrm{E}_{ij} \left( \frac{1}{r_i} - \frac{\beta_i}{\tau_i} \right) \\
 \label{eq:inti}  \tilde{w}^\mathrm{I}_{ik} &=& R^\mathrm{D}_i w^\mathrm{I}_{ik} \frac{\beta_i}{\tau_i}   \\
 \label{eq:xi}    \xi_i &=& R^\mathrm{D}_i \left(\frac{\beta_i}{\tau_i} - \gamma_i \right) \:.
\end{eqnarray}
The change in sign of the inhibitory contribution from \Eq{eq:lifneuron} to \Eq{eq:dfc2} has an obvious physical interpretation: In \pref{eq:lifneuron}, the change of membrane potential $U_i$ and therefore the spike rate is enhanced by EPSPs but diminished by IPSPs. On the other hand, the dendritic shunting current $I^\mathrm{D}_i$ in \pref{eq:dfc2} is large for both, large EPSPs and large IPSPs.

From \Eq{eq:lifneuron} we eventually obtain the neural network's dynamics by taking into account that postsynaptic potentials are obtained from presynaptic spike trains through temporal convolution with postsynaptic impulse response functions, i.e.
\begin{equation}
 \label{eq:psp}
  E^\mathrm{E|I}_{ij}(t) =  \int_{-\infty}^t s^\mathrm{E|I}_i(t - t') R_j(t') \; \Df t'
\end{equation}
where $s^\mathrm{E|I}_i(t)$ are excitatory and inhibitory synaptic impulse response functions, respectively, and $R_j$ is the spike train
\begin{equation}
 \label{eq:spiketrains}
 R_j(t) = \sum_{t_\nu} \delta(t - t_\nu - \tau_L)
\end{equation}
coming from presynaptic neuron $j$, when spikes were emitted at times $t_\nu$. The additional time constant $\tau_L$ is attributed to synaptic transmission delay \citep{MazzoniPanzeriEA08}. These events are obtained by integrating \pref{eq:lifneuron} with initial condition
\begin{equation}
\label{eq:spikeini}
 U_i(t_\nu) = E.
\end{equation}
Where $E$ is some steady-state potential~\citep{MazzoniPanzeriEA08}. If at time $t = t_{\nu}$ the membrane reaches a threshold
\begin{equation}
\label{eq:reset2}
 U_i(t) \ge \theta_i(t)
\end{equation}
(with possibly a time dependent activation threshold $\theta_i(t)$) from below $\frac{\Df U_i(t)}{\Df t} > 0$ then an output spike $\delta(t - t_{\nu})$ is generated, which is then followed by a potential resetting as follows
\begin{equation}
\label{eq:reset1}
 U_i(t_{\nu + 1}) \leftarrow E \:.
\end{equation}
Additionally, the integration of the dynamical law is restarted at time $t = t_{\nu + 1} + \tau_{rp}$ after interrupting the dynamics for a refractory period $\tau_{rp}$.

Inserting \pref{eq:psp} into \pref{eq:lifneuron} entails the evolution equation of the neural network
\begin{equation}
 \label{eq:lineuron2}
 \tau_i \frac{\Df U_i}{\Df t} + U_i =
 \sum_{j = 1}^p w^\mathrm{E}_{ij} \, s^\mathrm{E}_i(t) * R_j(t) +
 \sum_{k = 1}^q w^\mathrm{I}_{ik} \, s^\mathrm{I}_i(t) * R_k(t) \:,
\end{equation}
where the signs had been absorbed by the synaptic weights, such that $w^\mathrm{E}_{ij} > 0$ for excitatory synapses and $w^\mathrm{I}_{ik} < 0$ for inhibitory synapses, respectively.

Following \citet{MazzoniPanzeriEA08} an individual postsynaptic current $I^\mathrm{E|I}_{ij}$ at a synapse between neurons $i$ and $j$ obeys
\begin{eqnarray}
 \label{eq:pscdyn1}
  \tau^\mathrm{E|I}_d \frac{\Df I^\mathrm{E|I}_{ij}}{\Df t} + I^\mathrm{E|I}_{ij} &=& x^\mathrm{E|I}_{ij} \\
 \label{eq:pscdyn2}
  \tau^\mathrm{E|I}_r \frac{\Df x^\mathrm{E|I}_{ij}}{\Df t} + x^\mathrm{E|I}_{ij} &=& F^\mathrm{E|I}_{ij} \:,
\end{eqnarray}
where $\tau^\mathrm{E|I}_d$ are decay time constants and $\tau^\mathrm{E|I}_r$ are rise time constants of EPSC and IPSC, respectively. Auxiliary variables are denoted by $x^\mathrm{E|I}_{ij}$, while $F^\mathrm{E|I}_{ij}$ prescribes presynaptic forcing
\begin{equation}\label{eq:force}
    F^\mathrm{E|I}_{ij} = \tau_i J_{ij} R_j(t)
\end{equation}
with spike train \pref{eq:spiketrains}. Here, $J_{ij} = v w^\mathrm{E|I}_{ij}$ denotes synaptic gain with $v = \quant{1}{mV}$ as voltage unit.

Note that \pref{eq:force} is essentially a weighted sum of delta functions, such that a single spike can be assumed as particular forcing
\begin{equation}\label{eq:forcesing}
    F = F_0 \delta(t) \:,
\end{equation}
with some constant $F_0$.

Derivating \pref{eq:pscdyn1} and eliminating $x^\mathrm{E|I}_{ij}$ transforms Eqs. (\ref{eq:pscdyn1} -- \ref{eq:pscdyn2}) into a linear second-order differential equation with constant coefficients
\begin{equation}\label{eq:psc2ode}
 \tau^\mathrm{E|I}_d \tau^\mathrm{E|I}_r \frac{\Df^2 I^\mathrm{E|I}_{ij}}{\Df t^2} +
 (\tau^\mathrm{E|I}_d + \tau^\mathrm{E|I}_r) \frac{\Df I^\mathrm{E|I}_{ij}}{\Df t} +
 I^\mathrm{E|I}_{ij} = F^\mathrm{E|I}_{ij} \:.
\end{equation}

Equation \pref{eq:psc2ode} with the particular forcing \pref{eq:forcesing} is solved by a Green's function $s_i^\mathrm{E|I}(t)$ such that the general solution of \pref{eq:psc2ode} is obtained as the temporal convolution
\begin{equation}\label{eq:greenconvo}
 I^\mathrm{E|I}_{ij}(t) = \int_{-\infty}^t s_i^\mathrm{E|I}(t - t') F^\mathrm{E|I}_{ij}(t) \; \Df t' \:.
\end{equation}

For $t \ne 0$, \pref{eq:psc2ode} assumes its homogeneous form and is easily solved by means of the associated characteristic polynomial
\begin{equation}\label{eq:chapoly}
 \tau^\mathrm{E|I}_d \tau^\mathrm{E|I}_r \lambda^2 +
 (\tau^\mathrm{E|I}_d + \tau^\mathrm{E|I}_r) \lambda +
 1 = 0
\end{equation}
with roots $\lambda_1 = -1/\tau^\mathrm{E|I}_d$ and $\lambda_2 = -1/\tau^\mathrm{E|I}_r$, entailing the Green's functions
\begin{equation}\label{eq:greens}
   s_i^\mathrm{E|I}(t) = \left( A^\mathrm{E|I} \Ex^{t/\tau^\mathrm{E|I}_r}  -
   B^\mathrm{E|I}  \Ex^{t/\tau^\mathrm{E|I}_d} \right) \Theta(t)
\end{equation}
with the Heaviside step function $\Theta(t)$.

The constants $A^\mathrm{E|I}, B^\mathrm{E|I} > 0$ are obtained from the initial conditions $s_i^\mathrm{E|I}(t) = 0$, reflecting causality, and a suitable normalization
\[
  \int_0^\infty s_i^\mathrm{E|I}(t) \Df t = 1 \:.
\]

The initial condition yields $A^\mathrm{E|I} = B^\mathrm{E|I} \equiv S^\mathrm{E|I}$, while the remaining constant
\[
  S^\mathrm{E|I} = \frac{1}{\tau^\mathrm{E|I}_d - \tau^\mathrm{E|I}_r} \:,
\]
due to normalization. Therefore, the normalized Green's functions are those of \citet{BrunelWang03}
\begin{equation}\label{eq:greensnorm}
   s_i^\mathrm{E|I}(t) = v \frac{\tau_i}{\tau^\mathrm{E|I}_d - \tau^\mathrm{E|I}_r} \left( \Ex^{t/\tau^\mathrm{E|I}_r}  -
   \Ex^{t/\tau^\mathrm{E|I}_d} \right) \Theta(t) \:.
\end{equation}

Now, we are able to compare our DFP $V_i$ [\Eq{eq:dfc2}] with the estimate of \citet{MazzoniPanzeriEA08} which is given (in our notation) as the sums of the moduli of excitatory and inhibitory synaptic currents, i.e.
\begin{equation}
 \label{eq:mazodfp}
 V_i^\mathrm{MPLB} = \sum_j |I^\mathrm{E}_{ij}| + \sum_k |I^\mathrm{I}_{ik}|
\end{equation}
where ``MPLB'' refers to the authors \citet{MazzoniPanzeriEA08}.

From \pref{eq:dfc2} and \pref{eq:mazodfp}, respectively, we compute two models of the local field potential (LFP). First, by summing DFP across all pyramidal neurons \citep{GrabenKurths08, MazzoniPanzeriEA08}, and, second by taking the DFP average \citep{NunezSrinivasan06},  which yields
\begin{eqnarray}
    \label{eq:lfp1} L_1 &=& \sum_i V_i^\mathrm{MPLB} \\
    \label{eq:lfp2} L_2 &=& \frac{1}{K} \sum_i V_i^\mathrm{MPLB} \\
    \label{eq:lfp3} L_3 &=& \sum_i V_i \\
    \label{eq:lfp4} L_4 &=& \frac{1}{K} \sum_i V_i \:,
\end{eqnarray}
where $K$ is number of pyramidal neurons.

\subsection{Parameter estimation}
\label{sec:est}

Next, we relate the electrotonic parameters of our model to the phenomenological parameters of \citet{MazzoniPanzeriEA08}. To this end, we first report their synaptic efficacies in \Tab{tab:synapticeficacies}.

\begin{table}[H]
  \centering
  \begin{tabular}{lcc}
  Synaptic efficacies / mV & on interneurons & on pyramidal neurons\\
  \hline
  GABA &  2.7 & 1.7 \\
  recurrent cortical AMPA & 0.7 & 0.42\\
  external thalamic AMPA  &  0.95 &  0.55\\
  \end{tabular}
  \caption{Parameters laid as in~\citet{MazzoniPanzeriEA08}.}\label{tab:synapticeficacies}
\end{table}

From these, we compute the synaptic weights through
\begin{equation}
  \label{eq:SynapticWeight}
  w^\mathrm{E}_{ij} = J^\mathrm{E}_{ij} / v = \left\{
        \begin{array}{r@{\quad \text{if} \quad}l}
          0.42 & j \text{ ``cortical''} \\
          0.55 & j \text{ ``thalamic''}
        \end{array}
  \right.
\end{equation}
and
\[
  w^\mathrm{I}_{ik} = J^\mathrm{I}_{ik} / v = 1.7
\]

Next, we determine the factors $r_i$ by virtue of \Eq{eq:inweig} through
\[
  r_i = \frac{w^\mathrm{I}_{ik}}{\bar{g}_\mathrm{GABA}} = \frac{1.7}{\quant{1}{nS}} = \quant{1.7}{G\Omega}
\]
using the inhibitory synaptic conductivity $\bar{g}_\mathrm{GABA} = \quant{1}{nS}$, Correspondingly, \Eq{eq:exweig} allows us to express $\alpha_{ij}$ in terms of the excitatory synaptic weights through
\[
  \alpha_{ij} = \frac{w^\mathrm{E}_{ij}}{r_i} = \left\{
        \begin{array}{r@{\quad \text{if} \quad}l}
          \quant{0.25}{nS}  & j \text{ ``cortical''} \\ 
          \quant{0.32}{nS}  & j \text{ ``thalamic''}   
        \end{array}
  \right.
\]

From $\alpha_{ij}$ we can determine the total excitatory synaptic conductivities $g^\mathrm{E}_i$ according to \Eq{eq:alpha} through
\begin{eqnarray*}
  \alpha_{ij} &=& \frac{1}{R^\mathrm{E}_{ij}[ 1 + g^\mathrm{E}_i (R^\mathrm{A}_i + R^\mathrm{D}_i) ]} \\
  g^\mathrm{E}_i \left[ 1- (R^\mathrm{A}_i + R^\mathrm{D}_i) \sum_{j=1}^p \alpha_{ij} \right]  &=& \sum_{j=1}^p \alpha_{ij} \\
\end{eqnarray*}
\begin{equation}\label{eq:totexcond}
    g^\mathrm{E}_i = \frac{\sum_{j=1}^p \alpha_{ij}}
                                {1 - (R^\mathrm{A}_i + R^\mathrm{D}_i) \sum_{j=1}^p \alpha_{ij} }
\end{equation}
and hence
\begin{equation}\label{eq:exres}
  R^\mathrm{E}_{ij} = \frac{1}{\alpha_{ij} [ 1 + g^\mathrm{E}_i (R^\mathrm{A}_i + R^\mathrm{D}_i) ]}
\end{equation}

Inserting next \pref{eq:beta} into \pref{eq:timeconst} yields
\begin{eqnarray}\label{eq:tau2}
  \tau_i  &=&
      r_i C_i \frac{ 1 + g^\mathrm{E}_i (R^\mathrm{A}_i + R^\mathrm{D}_i) + (R^\mathrm{B}_i + R^\mathrm{C}_i) \{ g^\mathrm{E}_i - g^\mathrm{I}_i [ 1 + g^\mathrm{E}_i (R^\mathrm{A}_i + R^\mathrm{D}_i) ] \} }{1 + g^\mathrm{E}_i (R^\mathrm{A}_i + R^\mathrm{D}_i) }  \:.
\end{eqnarray}

Equation \pref{eq:tau2} could constraint the choice of the membrane capacitance $C_i$ by choosing $\tau_i = \quant{20}{ms}$ \citep{MazzoniPanzeriEA08}.

In order to also determine the DFP parameters  \pref{eq:exti} -- \pref{eq:xi}, we finally compute the ratios
\begin{eqnarray*}
  \frac{\beta_i}{\tau_i} &=&
  \frac{g^\mathrm{E}_i (R^\mathrm{B}_i + R^\mathrm{C}_i) }{ r_i \{ 1 + g^\mathrm{E}_i (R^\mathrm{A}_i + R^\mathrm{D}_i) + (R^\mathrm{B}_i + R^\mathrm{C}_i) \{ g^\mathrm{E}_i - g^\mathrm{I}_i [ 1 + g^\mathrm{E}_i (R^\mathrm{A}_i + R^\mathrm{D}_i) ] \} \} }
  \:.
\end{eqnarray*}

The remaining electrotonic parameters $R^\mathrm{M}_i$, $R^\mathrm{A}_i, R^\mathrm{B}_i, R^\mathrm{C}_i$, and $R^\mathrm{D}_i$ are estimated from cell geometries as follows. The resistance $R$ of a volume conductor is proportional to its length $\ell$ and reciprocally proportional to its cross-section $A$, i.e.
\begin{equation}\label{eq:resistor}
    R = \rho \frac{\ell}{A}
\end{equation}
where $\rho$ is the (specific) resistivity of the medium. Table \ref{tab:res} shows the resistivities of the three kinds of interest which then allows to evaluate the various volume conductor resistances according to \Eq{eq:resistor}.

\begin{table}[H]
  \centering
  \begin{tabular}{ll}
  medium & $\rho /  \mathrm{\Omega cm}$ \\
  \hline
  cell membrane (at axon hillock) & $5 \cdot 10^7$  \\
  cell plasma (cytoplasm) & 200 \\
  extracellular space & 333 \\
  \end{tabular}
  \caption{Resistivities of cell membrane, cell plasma and extracellular space. Parameters from \citet{kole2012signal, rall1977core, mainen1995model, gold2007using}. Note that the resistivity of the cell membrane has to be related to the constant membrane thickness ($\approx \quant{10}{nm}$). }\label{tab:res}
\end{table}

We consider a total dendritic length of $2 \ell = \quant{20}{\mu m}$ and a dendritic radius of $a = \quant{7}{\mu m}$, that are generally subjected to variation. Equally, parameters that were allowed to vary are the length and radius of the axon hillock, yet herein we consider a length of $2 \ell = \quant{20}{\mu m}$ and radius of $a = \quant{0.5}{\mu m}$ \citep{Destexhe01, kole2012signal, mainen1995model}. To evaluate the intracellular ($R_A$, $R_B$) and extracellular ($R_D$, $R_C$) resistances, respectively, according to \Eq{eq:resistor}, we consider a simple implementation where the length $\ell$ is half of the dendritic length (i.e. basal and apical length are symmetrical, but this can be broken). However, the cross sectional area for the cytoplasm is simply $A =\pi a^2$. Finally, the area of the axon hillock is simply the surface area of a cylinder.

In order to also determine the cross-section of extracellular space between dendritic trunks we make the following approximations. We assume that dendritic trunks are parallel aligned cylinders of radius $a$ and length $\ell$ that are hexagonally dense packed. Then the centers of three adjacent trunks form an equilateral triangle with side length $2a$ and hence area $2 \sqrt{3} a^2$. The enclosed space is then given by the difference of the triangle area and the area of three sixth circle sectors, therefore
\[
  A_\mathrm{space} = 2 \sqrt{3} a^2 - \frac{3}{6} \pi a^2 = \left( 2 \sqrt{3} - \frac{1}{2} \pi \right) a^2 \:.
\]

Hence, the cross-section of extracellular space surrounding one trunk is
\begin{equation}\label{eq:crossec}
    A = 6 A_\mathrm{space} = \left( 12 \sqrt{3} - 3 \pi \right) a^2 \:.
\end{equation}

\subsection{Simulations}
\label{sec:sim}

Subsequently, we implement an identical network to the one considered by \citet{MazzoniPanzeriEA08} with \emph{Brian Simulator}, that is a Python based environment~\citep{goodman2009brian}. However, the derivations from the previous section enables the possibility of setting a dipole observable that measures the local dendritic field potential (DFP) on each pyramidal neurons, given by \Eq{eq:dfc2}. This allows then to define a mesoscopic LFP observable, which can be equated either as averaged DFP or simply given as the sum of DFP, given by Eqs. (\ref{eq:lfp1} -- \ref{eq:lfp4}). Primarily, we compare our LFP measure $L_4$, proposed as the average of DFP, with the \citeauthor{MazzoniPanzeriEA08} LFP $L_1$ which is defined as the sum of absolute values of GABA and AMPA currents \Eq{eq:mazodfp}. Additionally, we also compare all possible measures, namely, mean membrane potential $\frac{1}{K}\sum_i U_i$, \citeauthor{MazzoniPanzeriEA08} LFP $L_1$,  average of \citeauthor{MazzoniPanzeriEA08} DFP $L_2$, sum of DFP $L_3$, and the average of DFP $L_4$.

For completeness, we briefly summarize the description of the network (we refer the reader to~\citet{MazzoniPanzeriEA08} for details). The network models a cortical tissue with leaky integrate-and-fire neurons, composed of 1000 inhibitory interneurons and 4000 pyramidal neurons, which are described by the evolution equation~(\ref{eq:lineuron2}). The threshold crossings given by \Eq{eq:reset2} is considered static with $\theta_i = \quant{18}{mV}$ and the reset potential $E=\quant{11}{mV}$. The refractory period for excitatory neurons is $\tau_{rp} = \quant{2}{ms}$ while for inhibitory neurons it is $\tau_{rp} =\quant{1}{ms}$. The network connectivity is random and sparse with a $0.2$ probability of directed connection between any pair of neurons. The evolution of synaptic currents, fast GABA (inhibitory) and AMPA (excitatory) are described via the second order evolution equations~(\ref{eq:pscdyn1} -- \ref{eq:pscdyn2}), which are activated by incoming presynaptic spikes represented by \Eq{eq:spiketrains}. The latency of the postsynaptic currents is set to $\tau_L=\quant{1}{ms}$ and the rise and decay times are given by \Tab{tab:synaptictimes}.

\begin{table}[H]
  \centering
  \begin{tabular}{lll}
  Synaptic times & $\tau_r$ / ms & $\tau_d$ / ms\\
  \hline
  GABA & 0.25 & 5 \\
  AMPA on interneurons & 0.2 & 1\\
  AMPA on pyramidal neurons  & 0.4 & 2\\
  \end{tabular}
  \caption{Synaptic rise ($\tau_r$) and decay times ($\tau_d$). Parameters laid as in~\citet{MazzoniPanzeriEA08}.}\label{tab:synaptictimes}
\end{table}

Moreover, synaptic efficacies, $J_{ij}^{E|I}$, for simulation were presented in \Tab{tab:synapticeficacies}. Note that relation~(\ref{eq:SynapticWeight}) then allows to determine the synaptic weights. Additionally, all neurons receive external thalamic excitatory inputs, that is, via AMPA-type synapses, which are activated by random Poisson spike trains, with a time varying rate that is identical for all neurons. Specifically, the thalamic inputs are the only source of noise, which attempts to account for both cortical heterogeneity and spontaneous activity. This is achieved by modeling a two level noise, where the first level is an Ornstein-Uhlenbeck process superimposed with a constant signal and the second level is a time varying inhomogeneous Poisson process. Thus, we have the following time varying rate, $\lambda(t)$, that feeds into inhomogeneous Poisson process:
\begin{eqnarray}
\tau_n \frac{\Df n(t)}{\Df t} &=& - n(t) + \sigma_n \sqrt{\frac{2}{\tau_n}} \, \eta(t)\\
\lambda(t) &=& [c_0 + n(t)]_{+}
\end{eqnarray}
where $\eta(t)$ represents Gaussian white noise, $c_0$ represents a constant signal (but equally could be periodic or other), and the operation $[\cdot]$ is the threshold-linear function, $[x]_{+}=x$ if $x > 0$, $[x]_{+}=0$ otherwise, which circumvents negative rates. The constant signal $c_0$ can range between $1.2$ to $2.6$ spikes/ms. The parameters of the Ornstein-Uhlenbeck process are $\tau_n=\quant{16}{ms}$ and the standard deviation $\sigma_n=0.4$ spikes/ms.

For complete exposition, we note that from an implementation viewpoint (within the Brian simulator), a copy of the postsynaptic impulse response function~(\ref{eq:psp}) has to be evaluated to calculate the dendritic field potential (DFP)~(\ref{eq:dfc2}) with weights $\tilde{w}^\mathrm{E|I}_{ij}$. This implies evaluating the second order process~(\ref{eq:pscdyn1}) -- (\ref{eq:pscdyn2}) with a different forcing term. Specifically, starting from $I^\mathrm{E|I}_{ij}(t) \equiv  w^\mathrm{E|I}_{ij} E^\mathrm{E|I}_{ij}(t) = s_i^\mathrm{E|I}(t) *F^\mathrm{E|I}_{ij}$ and pre-multiplying both sides with $\tilde{w}^\mathrm{E|I}_{ij}$ and subsequently re-arranging we obtain the desired forcing term $\tilde{F}^\mathrm{E|I}_{ij} = \tilde{w}^\mathrm{E|I}_{ij} F^\mathrm{E|I}_{ij}/w^\mathrm{E|I}_{ij}$. Note further that by expanding the term $F^\mathrm{E|I}_{ij}$ with equation~(\ref{eq:force}) and using relation~(\ref{eq:SynapticWeight}) we finally obtain $\tilde{F}^\mathrm{E|I}_{ij} = \tilde{w}^\mathrm{E|I}_{ij} \tau_i v R_j(t)$.

\section{Results}
\label{sec:res}

\begin{figure}[H]
 \centering
  \includegraphics[width=0.8\textwidth]{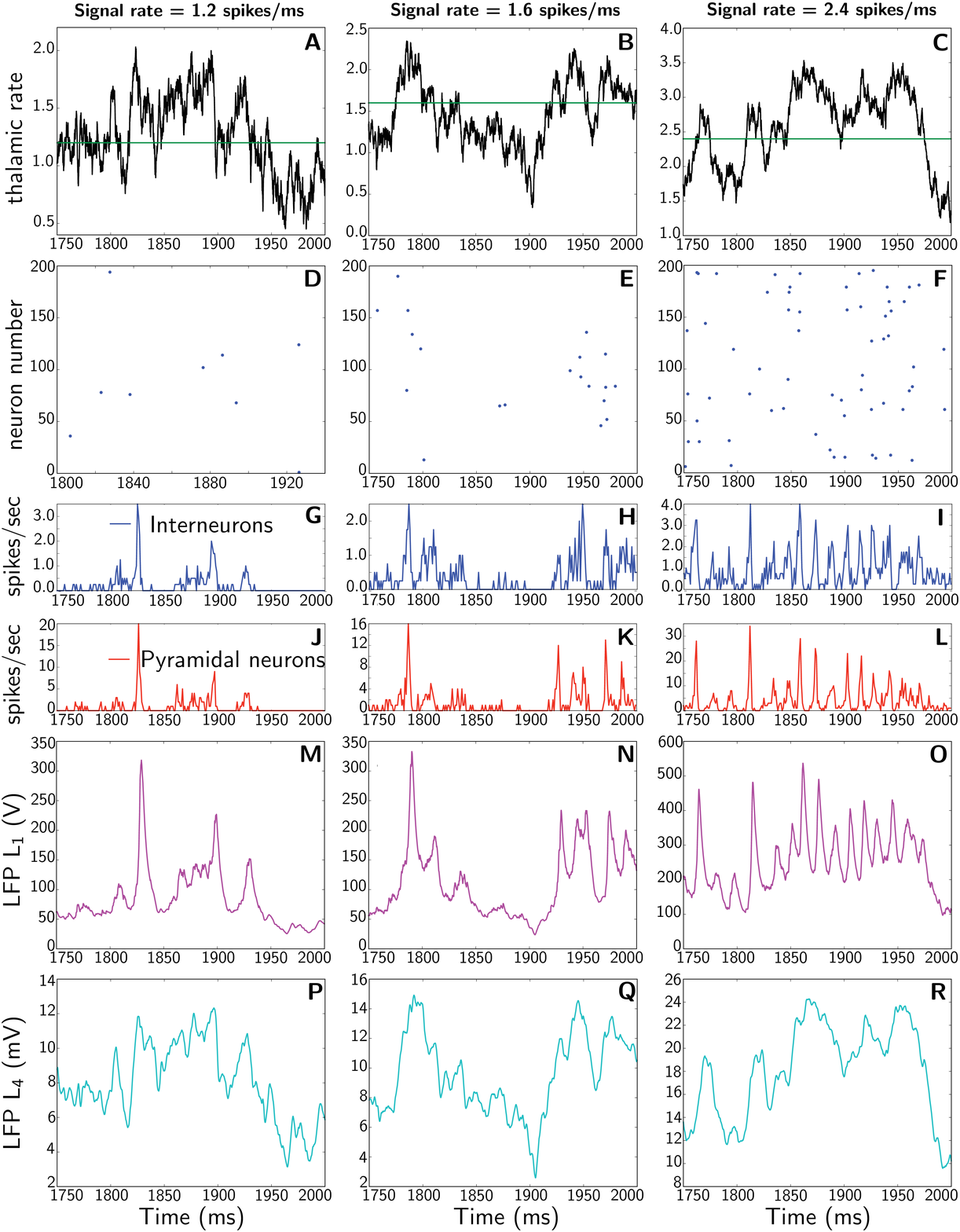}
\caption{\label{fig:lfp1} Dynamics of the network and LFP comparisons: The three columns represent different runs of the network for three different rates, 1.2, 1.6 and 2.4 spikes/ms. In each column, all panels show the same 250 ms (extracted from 2 seconds simulations). The top panels (A-C) represent thalamic inputs with the different rates. The second top panels (D-F) corresponds to a raster plot of the activity of 200 pyramidal neurons. Panels (G-I) depict average instantaneous firing rate (computed on a 1 ms bin) of interneurons (blue) and panels (J-L) correspond to average instantaneous firing rate of pyramidal neurons. Panels (M-O) show the \citeauthor{MazzoniPanzeriEA08} LFP $L_1$ from \Eq{eq:lfp1}. Finally, panels (P-R) depict our proposed LFP measure $L_4$, which is the average of dendritic field potential (DFP) [\Eq{eq:lfp4}].
}
\end{figure}

Following~\citet{MazzoniPanzeriEA08}, the network simulations are run for two seconds with three different noise levels, specifically, receiving a constant signal with three different rates 1.2, 1.6 and 2.4 spikes/ms as depicted in \Fig{fig:lfp1}. Note that these input rates do not mean that a single neuron fires at these high rates. Rather, it can be obtained from multiple neurons that jointly fire with slower, yet desynchronized, rates converging at the same postsynaptic cell. The Poisson process ensures that this is well represented.

The focus is to compare our proposed measure $L_4$, defined as mean of the dendritic field potential (DFP) [\Eq{eq:lfp4}], with the \citeauthor{MazzoniPanzeriEA08} LFP $L_1$ from \Eq{eq:lfp1}. In \Fig{fig:lfp1} one sees two main striking differences between the two measures, namely in frequency and in amplitude. Specifically, $L_1$ responds instantaneously to the spiking network activity by means of high frequency oscillations. Moreover, $L_1$ also exhibits a large amplitude. In contrast, our mean dendritic field potential $L_4$ measures comparably to experimental LFP, that is, in the order of millivolts, and although it responds to population activity, it has a relatively smoother response. Actually one can realize that our LFP estimate represents low-pass filtered thalamic input.

The physiological relevance of this is not yet clear in our work. However, recent work~\citep{PouletEA2012} shows that desynchronized cortical state during active behavior is driven by a centrally generated increase in thalamic action potential firing (i.e. thalamic firing controls cortical states). Thus, it seems that cortical synchronous activity is suppressed when thalamic input increases, thereby suggesting that cortical desynchronized states to be related to sensory processing. This work further quantifies these observations by applying Fast Fourier Transform (FFT) to cortical EEG and subsequently comparing with thalamic firing rate by means of Pearson correlation coefficient. Unfortunately they do not quantify the amount of thalamic oscillations contained within the cortical EEG.

\begin{figure}[H]
 \centering
\includegraphics[width=0.8\textwidth]{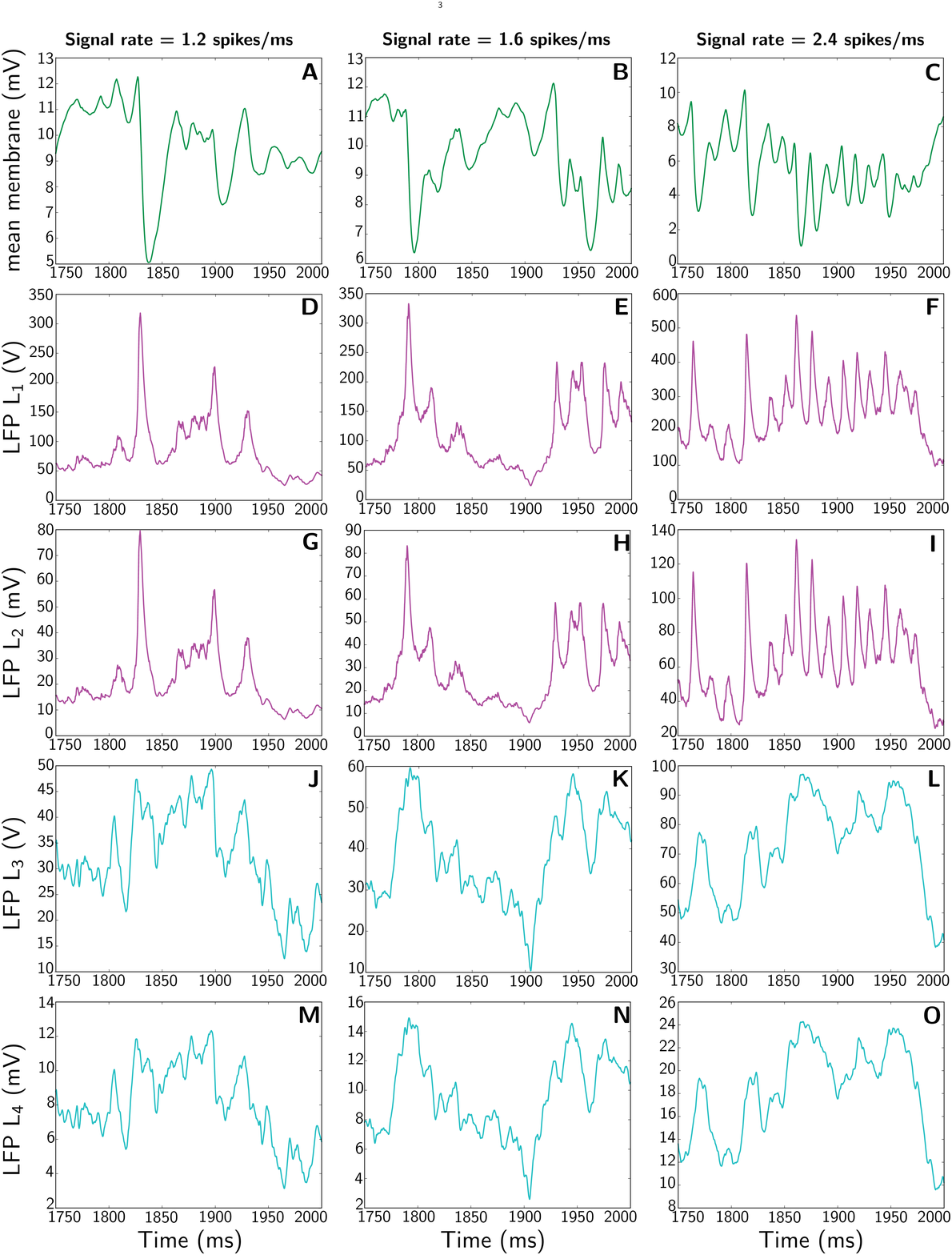}
\caption{\label{fig:lfp2} Comparison of different LFP measures when the network receives constant signal with three different rates (1.2, 1.6 and 2.4 spikes/ms). Again, only 250 ms is represented (extracted from 2 sec simulation). The first plot (A-C) corresponding to the the different rates shows the most widespread LFP  measure used in the literature, namely average membrane potential $\frac{1}{K}\sum_i U_i$. The second panel (D-F) shows the \citeauthor{MazzoniPanzeriEA08} LFP $L_1$ from \Eq{eq:lfp1}. The third panel (G-I) displays the average of the \citeauthor{MazzoniPanzeriEA08} DFP $L_2$ [\Eq{eq:lfp2}]. Similarly, the fourth panel (J-L) shows the total, $L_3$, [\Eq{eq:lfp3}]  and the last panel (M-O) depicts the averaged, $L_4$, [\Eq{eq:lfp4}] LFP measure. Note the different amplitude scales between measures.}
\end{figure}

Yet, to keep a comparable comparison between measures, we also compute the average of the \citeauthor{MazzoniPanzeriEA08} DFP $L_2$ [\Eq{eq:lfp4}] and additionally the mean membrane potential (the standard considered in the neuroscientific literature). These are shown in \Fig{fig:lfp2}.

Clearly, in terms of time profile, the summed and averaged observables are similar within the same class of LFP measures. However, in all cases the \citeauthor{MazzoniPanzeriEA08} LFP $L_1$  exhibits a significantly larger order of magnitude, which diverges substantially from experimental LFP amplitudes, typically varying between 0.5 to 2 mV \citep{niedermeyer20059,lakatos2005oscillatory}. In contrast, although the mean DFP is not contained within the interval from 0.5 to 2 mV it arguably performs better. However, we do concede further work is required. Some gains in improving the different LFP measures can be achieved by applying for example a weighted average, which would mimic the distance of an electrode to a particular neuron by means of a lead field kernel \citep{NunezSrinivasan06}. For example a convolution of either $L_1$ or $L_2$ with a Gaussian kernel (representing the distance to a neuron), would yield a measure that captures better the local field potential or better the dendritic field potential of the nearest neurons. However, further work will be required to properly quantify the gain when space is taken into account.

\begin{figure}[H]
 \centering
\includegraphics[width=0.8\textwidth]{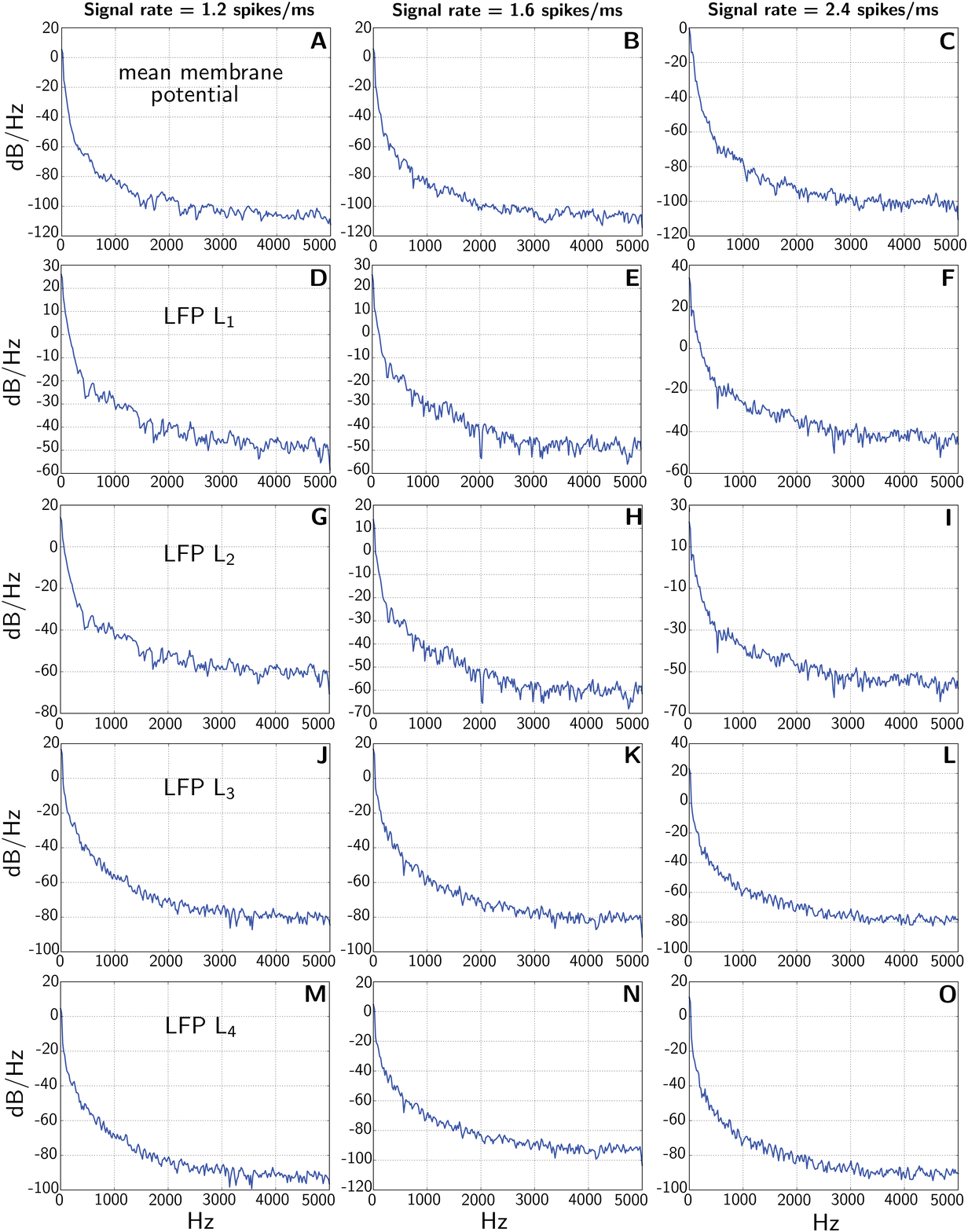}
\caption{\label{fig:lfp3} Comparison of power spectra of the various LFP measures when the network receives constant signal with three different rates (1.2, 1.6 and 2.4 spikes/ms): The first plot (A-C) corresponding to the different rates shows the power spectrum of the average membrane potential $\frac{1}{K}\sum_i U_i$. The second plot (D-F) and third plots (G-I) show power spectra of the total and average of $L_1$ and $L_2$ corresponding to \citet{MazzoniPanzeriEA08}, respectively. The fourth plot (J-L) and fifth plots (M-O) display power spectra of the $L_3$ and $L_4$ measures from our model, respectively. Note we show the full spectrum up to $\quant{5}{kHz}$ only for convenience due to the fine sample rate.}
\end{figure}

In \Fig{fig:lfp3} we finally contrast the power spectra of the different LFP measures.

One interesting feature is that the power spectrum of the \citeauthor{MazzoniPanzeriEA08} LFP measures decays much more slowly that the average membrane potential for higher frequencies. This observation is true for both, $L_1$ and $L_2$. In contrast, our LFP measures $L_3$ and $L_4$ fare better, and in particular, $L_4$ decays at an approximately similar rate as the average membrane potential.

\section{Discussion}
\label{sec:disc}

In this article we derived a model for cortical dipole fields, such as dendritic and local field potential (DFP/LFP) from biophysical principles. To that aim we decomposed a cortical pyramidal cell, the putative generator of those potentials, into three  compartments: the apical dendritic tree as the place of mainly excitatory (AMPA) synapses, the soma and the perisomatic dendritic tree as the place of mainly inhibitory (GABA) synapses and the axon hillock as the place of wave-to-spike conversion by means of an integrate-and-fire mechanism. From Kirchhoff's laws governing an electronic equivalent circuit of our model, we were then able to derive the evolution equation for neural network activity \Eq {eq:lineuron2} and, in addition, an observation equation \Eq{eq:dfc2} for the dendritic dipole potential contributing to the LFP of a cortical population.

In order to compare our approach with another model discussed in the recent literature \citep{MazzoniPanzeriEA08, MazzoniEA10, MazzoniBrunelEA11} we aligned the parameters of our model with the model of \citet{MazzoniPanzeriEA08} who approximated DFP as the sum of moduli of excitatory and inhibitory synaptic currents \Eq{eq:mazodfp}. From both approaches, we computed four different LFP estimates: $L_1$, the sum of \citeauthor{MazzoniPanzeriEA08} DFP,  $L_2$, the population average of \citeauthor{MazzoniPanzeriEA08} DFP,  $L_3$ the sum of our dipole DFP, and $L_4$ the population average of our dipole DFP [Eqs. (\ref{eq:lfp1} -- \ref{eq:lfp4})].

Our results indicate two main effects between our dipole LFP measures and those of \citeauthor{MazzoniPanzeriEA08} Firstly, the measures based on \citet{MazzoniPanzeriEA08} systematically overestimate LFP amplitude by almost one order of magnitude. One reason for that could be attributed to the direct conversion of synaptic current into voltage without taking extracellular conductivity into account, as properly done in our approach. Yet, another, even more crucial reason is disclosed by our equivalent circuit \Fig{fig:eqcirc}. In our approach there is just \emph{one} extracellular current $I^\mathrm{D}$ flowing from the perisomatic to the apical dendritic tree. In the model of \citet{MazzoniPanzeriEA08}, however, two synaptic currents that might be of the same order of magnitude are superimposed to the DFP. Secondly, the measures based on \citet{MazzoniPanzeriEA08} also systematically overestimate LFP frequencies. This could probably be attributed partly to spurious higher harmonics introduced by computing absolute values. Moreover, taking the power spectrum shows that the \citet{MazzoniPanzeriEA08} measure decays much more slowly than the average membrane potential, which is at variance with experimental data.

However, at the current stage, both models, that of \citet{MazzoniPanzeriEA08} and our own, agree with respect to the polarity of DFP and LFP. The measures based on \citet{MazzoniPanzeriEA08} have positive polarity simply due to the moduli. On the other hand, also the direction of current dipoles in our model is constrained by the construction of the equivalent circuit \Fig{fig:eqcirc} where current sources are situated at the perisomatic and current sinks are situated at apical dendritc tree. Taking this polarity as positive also entails positive DFP and LFP that could only change in strength. However, it is well known from brain anatomy that pyramidal cells appear in at least two layers, III and VI, of neocortex. This is reflected in experiments when an electrode traverses different layers by LFP polarity reversals, and, of course, by the fact that LFP and EEG oscillate between positive and negative polarity. Adapting our model to this situation could be straightforwardly accomplished in the framework of neural field theory by fully representing space and simulating layered neural fields \citep{Amari77b, JirsaHaken96b, Graben08a}. By contrast such a generalization is impossible at all with the model of \citet{MazzoniPanzeriEA08} due to the presence of absolute values.

On theses grounds we have good indication that our measure is an improvement to the \citeauthor{MazzoniPanzeriEA08} LFP measures, and, quite importantly, it is biophysically better motivated than the \emph{ad hoc} model of \citet{MazzoniPanzeriEA08}. However, much considerable effort is still required to underpin all the relevant LFP mechanisms and to better represent experimental LFP/EEG dynamics.

Finally, our work provides a new framework where dendritic field potentials and the relationship between firing rates and local fields can be explored without the extreme demand on computational complexity involved in  multicompartmental modeling \citep{LindenPettersenEinevoll10, LindenTetzlaffEA11, ProtopapasVanierBower98, SargsyanPapatheodoropoulosKostopoulos01} by adopting reduced compartment circuits. For example, we envisage to extend our recent work which maps firing rate model (derived from LIF models) to population density models \citep{Chizhov2007}, but now incorporating our observational DFP model. In addition, our framework is analytically amenable and thus can be applied to any linear differential equation, for instance, GIF (Gif-sur-Yvette Integrate Fire) models, which are improvements to the LIF models and compute more accurately spike activations \citep{rudolph2012analytical}. Also resonant membranes (mediated by $\mathrm{Ca}^{2+}$ and a $\mathrm{Ca}^{2+}$-activated $\mathrm{K}^{+}$ ionic currents) that describe sub-threshold oscillations and which can be easily expressed by linear equations~\citep{Mauro1970} can be incorporated in our derivations. We note however that our framework can be applied to non-linear equations, with \citet{HodgkinHuxley52} type activation, but it will fall short from explicit and analytical observation equations.

\section*{Acknowledgements}

We thank Michelle Lilith, Claude B\'{e}dard, Alain Destexhe, and J\"urgen Kurths for fruitful discussion. In addition, we would like to thank Samantha Adams for providing help with Brian Simulator installations and initial discussions of Brian usage. This research was supported by a DFG Heisenberg grant awarded to PbG (GR 3711/1-1).



\begin{thebibliography}{}

\bibitem[\protect\citeauthoryear{Amari}{Amari}{1977}]{Amari77b}
Amari, S.-I. (1977).
\newblock Dynamics of pattern formation in lateral-inhibition type neural
  fields.
\newblock {\em Biological Cybernetics\/}~{\em 27}, 77 -- 87.

\bibitem[\protect\citeauthoryear{B\'edard and Destexhe}{B\'edard and
  Destexhe}{2009}]{BedardDestexhe09}
B\'edard, C. and A.~Destexhe (2009).
\newblock Macroscopic models of local field potentials and the apparent $1/f$
  noise in brain activity.
\newblock {\em Biophysical Journal\/}~{\em 96\/}(7), 2589 -- 2603.

\bibitem[\protect\citeauthoryear{B\'{e}dard and Destexhe}{B\'{e}dard and
  Destexhe}{2012}]{BedardDestexhe12}
B\'{e}dard, C. and A.~Destexhe (2012).
\newblock Modeling local field potentials and their interaction with the
  extracellular medium.
\newblock In R.~Brette and A.~Destexhe (Eds.), {\em Handbook of Neural Activity
  Measurement}, pp.\  136 -- 191. Cambridge: Cambridge University
  Press.

\bibitem[\protect\citeauthoryear{B\'{e}dard, Kr\"{o}ger, and
  Destexhe}{B\'{e}dard et~al.}{2004}]{BedardKroegerDestexhe04}
B\'{e}dard, C., H.~Kr\"{o}ger, and A.~Destexhe (2004).
\newblock Modeling extracellular field potentials and the frequency-filtering
  properties of extracellular space.
\newblock {\em Biophysical Journal\/}~{\em 86\/}(3), 1829 -- 1842.

\bibitem[\protect\citeauthoryear{beim Graben}{beim Graben}{2008}]{Graben08a}
beim Graben, P. (2008).
\newblock Foundations of neurophysics.
\newblock In P.~b. Graben, C.~Zhou, M.~Thiel, and J.~Kurths (Eds.), {\em
  Lectures in Supercomputational Neuroscience: Dynamics in Complex Brain
  Networks}, Springer Complexity Series, pp.\  3 -- 48. Berlin:
  Springer.

\bibitem[\protect\citeauthoryear{beim Graben and Kurths}{beim Graben and
  Kurths}{2008}]{GrabenKurths08}
beim Graben, P. and J.~Kurths (2008).
\newblock Simulating global properties of electroencephalograms with minimal
  random neural networks.
\newblock {\em Neurocomputing\/}~{\em 71\/}(4), 999 -- 1007.

\bibitem[\protect\citeauthoryear{Berger}{Berger}{1929}]{Berger29}
Berger, H. (1929).
\newblock {\"Uber das Elektroenkephalogramm des Menschen}.
\newblock {\em Archiv f\"ur Psychiatrie\/}~{\em 87}, 527 -- 570.

\bibitem[\protect\citeauthoryear{Brunel and Wang}{Brunel and
  Wang}{2003}]{BrunelWang03}
Brunel, N. and X.-J. Wang (2003).
\newblock What determines the frequency of fast network oscillations with
  irregular neural discharges? {I.} synaptic dynamics and excitation-inhibition
  balance.
\newblock {\em Journal of Neurophysiology\/}~{\em 90\/}(1), 415 -- 430.

\bibitem[\protect\citeauthoryear{Chizhov, S., and R.}{Chizhov
  et~al.}{2007}]{Chizhov2007}
Chizhov, A.~V., R.~S., and T.~J. R. (2007).
\newblock A comparative analysis of a firing-rate model and conductance-based
  neural population model.
\newblock {\em Physics Letters A\/}~{\em 369}, 31 -- 36.

\bibitem[\protect\citeauthoryear{Creutzfeldt, Watanabe, and Lux}{Creutzfeldt
  et~al.}{1966a}]{CreutzfeldtWatanabeLux66a}
Creutzfeldt, O.~D., S.~Watanabe, and H.~D. Lux (1966a).
\newblock Relations between {EEG} phenomena and potentials of single cortical
  cells. {I.} evoked responses after thalamic and epicortical stimulation.
\newblock {\em Electroencephalography and Clinical Neurophysiology\/}~{\em
  20\/}(1), 1 -- 18.

\bibitem[\protect\citeauthoryear{Creutzfeldt, Watanabe, and Lux}{Creutzfeldt
  et~al.}{1966b}]{CreutzfeldtWatanabeLux66b}
Creutzfeldt, O.~D., S.~Watanabe, and H.~D. Lux (1966b).
\newblock Relations between {EEG} phenomena and potentials of single cortical
  cells. {II.} spontaneous and convulsoid activity.
\newblock {\em Electroencephalography and Clinical Neurophysiology\/}~{\em
  20\/}(1), 19 -- 37.

\bibitem[\protect\citeauthoryear{David and Friston}{David and
  Friston}{2003}]{DavidFriston03}
David, O. and K.~J. Friston (2003).
\newblock A neural mass model for {MEG/EEG}: coupling and neuronal dynamics.
\newblock {\em NeuroImage\/}~{\em 20}, 1743 -- 1755.

\bibitem[\protect\citeauthoryear{Destexhe}{Destexhe}{2001}]{Destexhe01}
Destexhe, A. (2001).
\newblock Simplified models of neocortical pyramidal cells preserving
  somatodendritic voltage attenuation.
\newblock {\em Neurocomputing\/}~{\em 38-40}, 167 -- 173.

\bibitem[\protect\citeauthoryear{Destexhe, Mainen, and Sejnowski}{Destexhe
  et~al.}{1998}]{DestexheMainenSejnowski98}
Destexhe, A., F.~Mainen, and T.~J. Sejnowski (1998).
\newblock Kinetic models of synaptic transmission.
\newblock See \citeN{KochSegev98}, pp.\  1 -- 25.

\bibitem[\protect\citeauthoryear{Gold, Henze, and Koch}{Gold
  et~al.}{2007}]{gold2007using}
Gold, C., D.~Henze, and C.~Koch (2007).
\newblock Using extracellular action potential recordings to constrain
  compartmental models.
\newblock {\em Journal of Computational Neuroscience\/}~{\em 23\/}(1), 39--58.

\bibitem[\protect\citeauthoryear{Goodman and Brette}{Goodman and
  Brette}{2009}]{goodman2009brian}
Goodman, D. and R.~Brette (2009).
\newblock The Brian simulator.
\newblock {\em Frontiers in Neuroscience\/}~{\em 3\/}(2), 192 -- 197.

\bibitem[\protect\citeauthoryear{Hodgkin and Huxley}{Hodgkin and
  Huxley}{1952}]{HodgkinHuxley52}
Hodgkin, A.~L. and A.~F. Huxley (1952).
\newblock A quantitative description of membrane current and its application to
  conduction and excitation in nerve.
\newblock {\em Journal of Physiology\/}~{\em 117}, 500 -- 544.

\bibitem[\protect\citeauthoryear{Holt and Koch}{Holt and
  Koch}{1999}]{HoltKoch99a}
Holt, G.~R. and C.~Koch (1999).
\newblock Electrical interactions via the extracellular potential near cell
  bodies.
\newblock {\em Journal of Computational Neuroscience\/}~{\em 6}, 169 -- 184.

\bibitem[\protect\citeauthoryear{Jansen and Rit}{Jansen and
  Rit}{1995}]{JansenRit95}
Jansen, B.~H. and V.~G. Rit (1995).
\newblock Electroencephalogram and visual evoked potential generation in a
  mathematical model of coupled cortical columns.
\newblock {\em Biological Cybernetics\/}~{\em 73}, 357 -- 366.

\bibitem[\protect\citeauthoryear{Jirsa and Haken}{Jirsa and
  Haken}{1996}]{JirsaHaken96b}
Jirsa, V.~K. and H.~Haken (1996).
\newblock Field theory of electromagnetic brain activity.
\newblock {\em Physical Review Letters\/}~{\em 77\/}(5), 960 -- 963.

\bibitem[\protect\citeauthoryear{Johnston and Wu}{Johnston and
  Wu}{1997}]{JohnstonWu97}
Johnston, D. and S.~M.-S. Wu (1997).
\newblock {\em Foundations of Cellular Neurophysiology}.
\newblock Cambridge (MA): MIT Press.

\bibitem[\protect\citeauthoryear{Koch and Segev}{Koch and
  Segev}{1998}]{KochSegev98}
Koch, C. and I.~Segev (Eds.) (1998).
\newblock {\em Methods in Neuronal Modelling. From Ions to Networks\/} (2nd
  ed.).
\newblock Computational Neuroscience. Cambridge (MA): MIT Press.

\bibitem[\protect\citeauthoryear{Kole and Stuart}{Kole and
  Stuart}{2012}]{kole2012signal}
Kole, M. and G.~Stuart (2012).
\newblock Signal processing in the axon initial segment.
\newblock {\em Neuron\/}~{\em 73\/}(2), 235--247.

\bibitem[\protect\citeauthoryear{Lakatos, Shah, Knuth, Ulbert, Karmos, and
  Schroeder}{Lakatos et~al.}{2005}]{lakatos2005oscillatory}
Lakatos, P., A.~Shah, K.~Knuth, I.~Ulbert, G.~Karmos, and C.~Schroeder (2005).
\newblock An oscillatory hierarchy controlling neuronal excitability and
  stimulus processing in the auditory cortex.
\newblock {\em Journal of Neurophysiology\/}~{\em 94\/}(3), 1904--1911.

\bibitem[\protect\citeauthoryear{Lind\'{e}n, Pettersen, and
  Einevoll}{Lind\'{e}n et~al.}{2010}]{LindenPettersenEinevoll10}
Lind\'{e}n, H., K.~Pettersen, and G.~Einevoll (2010).
\newblock Intrinsic dendritic filtering gives low-pass power spectra of local
  field potentials.
\newblock {\em Journal of Computational Neuroscience\/}~{\em 29}, 423 -- 444.

\bibitem[\protect\citeauthoryear{Lind\'en, T, Potjans, Pettersen, Gr\"un,
  Diesmann, and Einevoll}{Lind\'en et~al.}{2011}]{LindenTetzlaffEA11}
Lind\'en, H., T.~T, T.~C. Potjans, K.~H. Pettersen, S.~Gr\"un, M.~Diesmann, and
  G.~T. Einevoll (2011).
\newblock Modeling the spatial reach of the {LFP}.
\newblock {\em Neuron\/}~{\em 72\/}(5), 859 -- 872.

\bibitem[\protect\citeauthoryear{Mainen, Joerges, Huguenard, and
  Sejnowski}{Mainen et~al.}{1995}]{mainen1995model}
Mainen, Z., J.~Joerges, J.~Huguenard, and T.~Sejnowski (1995).
\newblock A model of spike initiation in neocortical pyramidal neurons.
\newblock {\em Neuron\/}~{\em 15\/}(6), 1427--1439.

\bibitem[\protect\citeauthoryear{Mauro, Conti, Dodge, and
  Schor}{Mauro et~al.}{1970}]{Mauro1970}
Mauro, A.,F.~Conti, F., F.~Dodge, and R.~Schor (1970).
\newblock Subthreshold behavior and phenomenological impedance of the squid giant axon.
\newblock {\em Journal of General Physiology\/}~{\em 55\/}, 497--532.

\bibitem[\protect\citeauthoryear{Mazzoni, Brunel, Cavallari, Logothetis, and
  Panzeri}{Mazzoni et~al.}{2011}]{MazzoniBrunelEA11}
Mazzoni, A., N.~Brunel, S.~Cavallari, N.~K. Logothetis, and S.~Panzeri (2011).
\newblock Cortical dynamics during naturalistic sensory stimulations:
  Experiments and models.
\newblock {\em Journal of Physiology\/}~{\em 105\/}(1-3), 2 -- 15.

\bibitem[\protect\citeauthoryear{Mazzoni, Panzeri, Logothetis, and
  Brunel}{Mazzoni et~al.}{2008}]{MazzoniPanzeriEA08}
Mazzoni, A., S.~Panzeri, N.~K. Logothetis, and N.~Brunel (2008).
\newblock Encoding of naturalistic stimuli by local field potential spectra in
  networks of excitatory and inhibitory neurons.
\newblock {\em PLoS Computational Biology\/}~{\em 4\/}(12), e1000239.

\bibitem[\protect\citeauthoryear{Mazzoni, Whittingstall, Brunel, Logothetis,
  and Panzeri}{Mazzoni et~al.}{2010}]{MazzoniEA10}
Mazzoni, A., K.~Whittingstall, N.~Brunel, N.~K. Logothetis, and S.~Panzeri
  (2010).
\newblock Understanding the relationships between spike rate and delta/gamma
  frequency bands of {LFPs} and {EEGs} using a local cortical network model.
\newblock {\em NeuroImage\/}~{\em 52\/}(3), 956 -- 972.

\bibitem[\protect\citeauthoryear{Niedermeyer}{Niedermeyer}{2005}]{niedermeyer20059}
Niedermeyer, E. (2005).
\newblock The normal EEG of the waking adult.
\newblock In E.~Niedermeyer and F.~L.~D. Silva (Eds.), {\em
  Electroencephalography: Basic principles, clinical applications, and related
  fields, 5th edition}, pp.\  167--192. Lippincott Williams \& Wilkins.

\bibitem[\protect\citeauthoryear{Nunez and Srinivasan}{Nunez and
  Srinivasan}{2006}]{NunezSrinivasan06}
Nunez, P.~L. and R.~Srinivasan (2006).
\newblock {\em Electric Fields of the Brain: The Neurophysics of {EEG}\/} (2nd
  ed.).
\newblock New York: Oxford University Press.

\bibitem[\protect\citeauthoryear{Omurtag, Knight, and Sirovich}{Omurtag
  et~al.}{2000}]{omurtag2000simulation}
Omurtag, A., B.~Knight, and L.~Sirovich (2000).
\newblock On the simulation of large populations of neurons.
\newblock {\em Journal of Computational Neuroscience\/}~{\em 8\/}(1), 51--63.

\bibitem[\protect\citeauthoryear{Poulet, Fernandez, Crochet,
  and Petersen}{Poulet et~al.}{2012}]{PouletEA2012}
Poulet, J.,F.A., L.M.J.~Fernandez, S.~Crochet, and C.H.~Petersen
  (2012).
\newblock Thalamic control of cortical states.
\newblock {\em Nature Neuroscience\/}~{\em 15\/}(3), 370 -- 372.

\bibitem[\protect\citeauthoryear{Protopapas, Vanier, and Bower}{Protopapas
  et~al.}{1998}]{ProtopapasVanierBower98}
Protopapas, A., M.~Vanier, and J.~M. Bower (1998).
\newblock Simulating large networks of neurons.
\newblock See \citeN{KochSegev98}, pp.\  461 -- 498.

\bibitem[\protect\citeauthoryear{Rall}{Rall}{1977}]{rall1977core}
Rall, W. (1977).
\newblock Core conductor theory and cable properties of neurons.
\newblock In E.~R. Kandel (Ed.), {\em Handbook of Physiology - The Nervous
  System, Cellular Biology of Neurons}, Volume~1, pp.\  39 -- 97. American
  Physiological Society.

\bibitem[\protect\citeauthoryear{Rodrigues, Chizhov, Marten, and
  Terry}{Rodrigues et~al.}{2010}]{rodrigues2010mappings}
Rodrigues, S., A.~Chizhov, F.~Marten, and J.~Terry (2010).
\newblock Mappings between a macroscopic neural-mass model and a reduced
  conductance-based model.
\newblock {\em Biological Cybernetics\/}~{\em 102\/}(5), 361--371.

\bibitem[\protect\citeauthoryear{Rudolph-Lilith, Dubois, and
  Destexhe}{Rudolph-Lilith et~al.}{2012}]{rudolph2012analytical}
Rudolph-Lilith, M., M.~Dubois, and A.~Destexhe (2012).
\newblock Analytical integrate-and-fire neuron models with conductance-based
  dynamics and realistic postsynaptic potential time course for event-driven
  simulation strategies.
\newblock {\em Neural Computation\/}~{\em 34}, 1426--1461.

\bibitem[\protect\citeauthoryear{Sargsyan, Papatheodoropoulos, and
  Kostopoulos}{Sargsyan et~al.}{2001}]{SargsyanPapatheodoropoulosKostopoulos01}
Sargsyan, A.~R., C.~Papatheodoropoulos, and G.~K. Kostopoulos (2001).
\newblock Modeling of evoked field potentials in hippocampal {CA1} area
  describes their dependence on {NMDA} and {GABA} receptors.
\newblock {\em Journal of Neuroscience Methods\/}~{\em 104}, 143 -- 153.

\bibitem[\protect\citeauthoryear{Schomer and Lopes~da Silva}{Schomer and
  Lopes~da Silva}{2011}]{SchomerSilva11}
Schomer, D.~L. and F.~H. Lopes~da Silva (Eds.) (2011).
\newblock {\em Niedermayer's Electroencephalography. Basic Principles, Clinical
  Applications, and Related Fields\/} (6th ed.).
\newblock Philadelphia: Lippincott Williams and Wilkins.

\bibitem[\protect\citeauthoryear{Spruston}{Spruston}{2008}]{Spruston08}
Spruston, N. (2008).
\newblock Pyramidal neurons: dendritic structure and synaptic integration.
\newblock {\em Nature Reviews Neuroscience\/}~{\em 9}, 206 -- 221.

\bibitem[\protect\citeauthoryear{Wang, Tegn\'er, Constantinidis, and
  Goldman-Rakic}{Wang et~al.}{2004}]{WangTegnerEA04}
Wang, X.-J., J.~Tegn\'er, C.~Constantinidis, and P.~S. Goldman-Rakic (2004).
\newblock Division of labor among distinct subtypes of inhibitory neurons in a
  cortical microcircuit of working memory.
\newblock {\em Proceedings of the National Academy of Sciences of the
  U.S.A.\/}~{\em 101\/}(5), 1368 -- 1373.

\bibitem[\protect\citeauthoryear{Wendling, Bellanger, Bartolomei, and
  Chauvel}{Wendling et~al.}{2000}]{WendlingBellangerEA00}
Wendling, F., J.~J. Bellanger, F.~Bartolomei, and P.~Chauvel (2000).
\newblock Relevance of nonlinear lumped-parameter models in the analysis of
  depth-{EEG} epileptic signals.
\newblock {\em Biological Cybernetics\/}~{\em 83}, 367 -- 378.

\bibitem[\protect\citeauthoryear{Wilson and Cowan}{Wilson and
  Cowan}{1972}]{WilsonCowan72}
Wilson, H.~R. and J.~D. Cowan (1972).
\newblock Excitatory and inhibitory interactions in localized populations of
  model neurons.
\newblock {\em Biophysical Journal\/}~{\em 12\/}(1), 1 -- 24.

\end{thebibliography}


\end{document}